\documentclass[aps,prd,nofootinbib,twocolumn]{revtex4}
\usepackage{graphicx}
\usepackage{amsmath,amssymb}
\usepackage{datetime}
\usepackage{helvet}
\usepackage{pstricks}
\usepackage{graphicx,color}
\usepackage{amssymb,amsmath}
\usepackage[colorlinks,bookmarksopen,bookmarksnumbered,citecolor=darkred, linkcolor=darkred,pdfstartview=FitH,urlcolor=darkred]{hyperref}
\newlength{\bibitemsep}
\newlength{\bibparskip}
\let\oldthebibliography\thebibliography
\renewcommand\thebibliography[1]{\oldthebibliography{#1}
 \setlength{\parskip}{\bibitemsep}
 \setlength{\itemsep}{\bibparskip}}
\definecolor{darkred}{rgb}{0.6471, 0.1098, 0.1882}
\definecolor{linkcolor}{rgb}{0.6471, 0.1098, 0.1882}
\definecolor{darkblue}{rgb}{0.1098, 0.1882, 0.6471}
\def\beqn#1{\begin{equation}\label{#1}}
\def\eeqn{\end{equation}}
\def\beqa#1{\begin{eqnarray}\label{#1}}
\def\eeqa{\end{eqnarray}}
\newcommand{\eq}[1]{eq.~(\ref{eq:#1})}

\newcommand{\secs}[1]{section~(\ref{sec:#1})}
\newcommand{\fig}[1]{fig.~(\ref{fig:#1})}
\newcommand{\nn}{\nonumber}

\def\z4{{\mathbb{Z}}_{4}}
\newcommand{\order}[1]{\mathcal{O}(#1)}
\newcommand{\gev}{\,\mathrm{GeV}}
\newcommand{\tev}{\,\mathrm{TeV}}
\newcommand{\kev}{\,\mathrm{keV}}

\begin{document}

\title{Dark Radiative Inverse Seesaw Mechanism}
\author{Amine Ahriche}
\email{aahriche@ictp.it} \affiliation{The Abdus Salam International
Centre for Theoretical Physics, Strada Costiera 11, I-34014,
Trieste, Italy;} \affiliation{Laboratory of Mathematical and
Sub-Atomic Physics (LPMPS), University of Constantine I, DZ-25000
Constantine, Algeria;} \affiliation{Department of Physics and Center
for Theoretical Sciences, National Taiwan University, Taipei 106,
Taiwan}

\author{Sofiane M. Boucenna}
\email{boucenna@lnf.infn.it}
\affiliation{INFN, Laboratori Nazionali di Frascati, C.P. 13, 100044 Frascati, Italy.}

\author{Salah Nasri}
\email{snasri@uaeu.ac.ae} \affiliation{Physics Department, UAE
University, POB 17551, Al Ain, United Arab Emirates}

\begin{abstract}
\noindent We present a minimal model that simultaneously accounts
for neutrino masses and the origin of dark matter (DM) and where the
electroweak phase transition is strong enough to allow for
electroweak baryogenesis. The Standard Model is enlarged with a
Majorana fermion, three generations of chiral fermion pairs, and a
single complex scalar that plays a central role in DM production and
phenomenology, neutrino masses, and the strength of the phase
transition. All the new fields are singlets under the SM gauge
group. Neutrino masses are generated via a new variant of  radiative
inverse seesaw where the required small mass term is generated via
loops involving DM and no large hierarchy is assumed among the mass
scales. The model offers all the advantage of low-scale neutrino
mass models as well as a viable dark matter candidate that is
testable with direct detection experiments.
\end{abstract}

\maketitle

\section{Introduction}

There are at least three concrete evidences which suggest that the
Standard Model (SM) is incomplete. These are $i$) non-zero neutrino
masses, $ii$) the existence of dark matter (DM), and $iii$) the
observation of the matter-antimatter asymmetry of the universe.
Neutrinos are usually assumed to be Majorana particles, in which
case an understanding of the origin of their mass necessarily
requires new degrees of freedom above the electroweak scale.
Similarly, the explanation for the $27\%$~\cite{Ade:2015xua} of the
total energy density of the universe in the form of DM implies the
need to extend the SM with at least one additional neutral particle that is
stable on cosmological time scales. The reason for the fact that the
observable universe is made of matter and not antimatter and that
the value of the cosmic baryon-to-photon ratio (i.e, the baryon asymmetry of the Universe (BAU)) is about $6\times
10^{-10}$ have no explanation within the SM either and require new physics.
While many extensions of the SM exist to solve these problems individually,
minimality as  dictated  by Occam's  razor  would  suggest  that models
offering simultaneous explanations to neutrinos and dark matter~\cite{Boucenna:2014zba,Restrepo:2013aga,Lattanzi:2014mia},
or dark matter and BAU ~\cite{Boucenna:2013wba,Davoudiasl:2012uw,Petraki:2013wwa,Zurek:2013wia}, or the three at once are favored. It is our goal here to address simultaneously all these shortcomings of the SM in a unified framework at the TeV scale.\\

The simplest mechanism for generating small neutrino masses is the
seesaw (type-I) mechanism
\cite{Minkowski:1977sc,Yanagida:1979as,Mohapatra:1979ia,GellMann:1980vs,Schechter:1980gr}
where three massive right-handed neutrinos are coupled to the left
handed neutrinos. However on the basis of naturalness, it invokes a
right handed neutrino with mass of order of Grand Unified Theories
(GUT) scale, making it hopeless to probe it in high energy physics
experiments. One way to lower the scale of the new physics is by
invoking `low-scale mechanisms'~\cite{Boucenna:2014zba}, in
particular the inverse seesaw where one extends the seesaw
mechanism with additional singlet fermions, $N_L$, and arrange for
the lepton charges such that the 2-units violation of lepton number
resides in the singlet  mass term $\mu N_L
N_L$~\cite{Mohapatra:1986bd}. The resulting light neutrino masses
are linearly proportional to $\mu$, $m_\nu \sim (m_D/M)^2 \mu $,
with $m_D$ and $M$ the usual Dirac mass and New Physics (NP) scale
respectively. It is clear then that if one chooses $\mu \sim
\text{keV}$, the scale of NP can be of order $\text{TeV}$. However,
the smallness of $\mu $ remains unexplained although
usually justified in terms of 't Hooft naturalness.\\

On the other hand, neutrino masses could be generated radiatively at a certain $n$-loop level.
The idea is that their mass can be naturally
small due to the loop suppression factor, $1/(16\, \pi^2)^{n}$, and the product of Yukawa
couplings instead of a suppression by the NP scale
\cite{Cheng:1980qt,Zee:1980ai, Ma:1998dn,Zee:1985id,Babu:1988ki,Krauss:2002px,Aoki:2008av, Gustafsson:2012vj}
 (for a review see \cite{Boucenna:2014zba}). This suppression allows the mass of the
new particles involved in the generation of neutrino masses to be
much smaller than the canonical seesaw mass scale. For instance, in
the three-loop neutrino mass generation models, the scale of the new
particles can be in the hundreds GeV scale, which makes them
testable at collider experiments \cite{Arhrib:2015dez, Babu:2002uu,
Ahriche:2014xra, Aoki:2009vf}. Furthermore, the use of discrete
symmetry that precludes the tree-level mass term for neutrinos,
allows the existence of DM candidate\footnote{A generalization of
\cite{Krauss:2002px} with septuplet representations
\cite{Ahriche:2015wha} has the interesting feature of automatically
containing stable DM candidate, without requiring a new discrete
symmetry.} which plays a role in the radiative neutrino mass
generation \cite{Ma:2006km,Ahriche:2014cda}, and could also trigger
the electroweak symmetry breaking \cite{ Ahriche:2014oda,
Ahriche:2015loa}.\\

In this work we propose a simple radiative inverse seesaw model where we extend the SM with
three chiral fermions and one complex scalar field that are all singlet under the SM gauge group.
A $\z4$ symmetry is invoked to simultaneously forbid the tree level inverse seesaw contribution
and provide a stable DM candidate.
The $\mu $ term is induced radiatively via
DM particles circulating in a loop. In this model
all the exotic particles have masses of order TeV scale or less,
which makes them accessible for collider experiments.  The
observed DM relic density can be naturally  obtained and the
spin-independent scattering cross section of the DM off nucleus is
consistent with the experimental limit reported by LUX
\cite{Akerib:2015rjg}, and yet within the reach of future DM direct
detection searches. In addition, a strongly first order electroweak
phase transition can be achieved, which is required for a successful
implementation of electroweak baryogenesis~\cite{Trodden:1998ym}.
We refer to Refs.~\cite{Ma:2009gu, Law:2012mj, Okada:2012np, Guo:2012ne,Baldes:2013eva,Fraser:2014yha,Huang:2014bva} for other radiative or linear inverse seesaw constructions.\\

This paper is organized as follows. In \secs{model} we present the
model. The generation of neutrino mass is presented in \secs{nu}.
In \secs{ewpt} we study the phenomenology of the scalar sector and
the strength of the electroweak phase transition. The calculation of
the DM relic abundance and direct detection is discussed in \secs{dm}.
Finally, we give our conclusion in \secs{end}.

\section{The model\label{sec:model}}

We consider a simple extension of the SM by adding three generations~%
\footnote{%
For simplicity we add the iso-singlet pairs sequentially, though two pairs
would suffice to account for the neutrino oscillations data.} of chiral
fermion pairs $N_{R}$ and $N_{L}$, one other chiral fermion $\chi _{R}\equiv
\chi $ and a complex scalar $S$. All the new fields transform trivially
under the SM gauge group, however we assign different charges to the fields of
the model under an imposed $\z4$ symmetry (or similarly, different $B-L$
 charges), c.f. table~(\ref{tab:model}). The SM quark sector is left
unchanged.\\

\begin{table}[t]
\centering
\begin{tabular}{l|c|c|ccc}
\hline & $L,\ell_{R},N_{R},N_{L}$ & $\chi$ & $H$ & $S$ & \\
\hline
$\z4$ & $i$ & -1 & +1 & $i$ & \\ \hline
\end{tabular}%
\caption{Summary of the relevant fields of the model and their quantum numbers.}
\label{tab:model}
\end{table}

The relevant terms in the Yukawa Lagrangian are the following (flavor indices are omitted):
\begin{eqnarray}
-\mathcal{L}&\supset& y_{\nu}\overline{L}\tilde{H}\,N_{R}+M\overline{N_{L}}N_{R}\nn\\
&&+\, y_{N}S\,\overline{\chi }\,N_{L}+\frac{m_{\chi }}{2}\chi ^{T}\,C^{-1}\chi +\mathrm{h.c.}\,, \label{eq:lag}
\end{eqnarray}
where $\tilde{H}\equiv i\sigma _{2}H^{\star }$ and $C$ is
the charge conjugation operator. The scalar potential is given by:
\begin{eqnarray}
V &=&-\mu _{H}^{2}\,H^{\dag }H+\tfrac{1}{2}\lambda _{H}\,(H^{\dag }H)^{2}
\notag \label{eq:pot} \\
&&+\mu _{S}^{2}S^{\star }S+\frac{\mu _{\nu }^{2}}{2}(S^{2}+\mathrm{h.c.})+%
\frac{\lambda _{S}}{2}(S^{\star }S)^{2}\nn\\
&&+\lambda _{HS}H^{\dagger }HS^{\star
}S.
\end{eqnarray}

The term $\mu _{\nu }^{2}$ which breaks the $\z4$ symmetry softly
is required by neutrino masses, as it is the origin of lepton number
violation (by two units). This will become clear in \secs{nu}. We see this term as a low energy manifestation of an
ultra-violet completion of the model and we remain agnostic as to
its specific origin. This could be for instance, a result of a
hidden sector that couples to the visible sector via the
`super-renormalizable' terms of the singlet $S$~\cite{Patt:2006fw}.
The mixed quartic coupling $\lambda_{HS}$ has to be positive because we found that negative
values destabilize the potential not far from the EW scale. With positive $\lambda_{HS}$ we always found stability up to at least
$10^6$ GeV, where the model is completed by a more complete theory.
In the next section
we address the neutrino phenomenology of the model.

\section{Neutrino masses \label{sec:nu}}

\begin{figure}[t!]
\begin{center}
\includegraphics[width=0.4\textwidth]{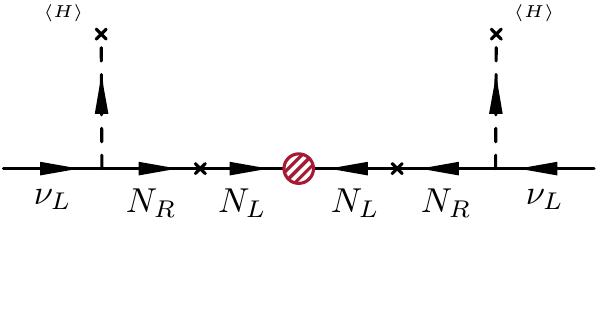}\\
\includegraphics[width=0.4\textwidth]{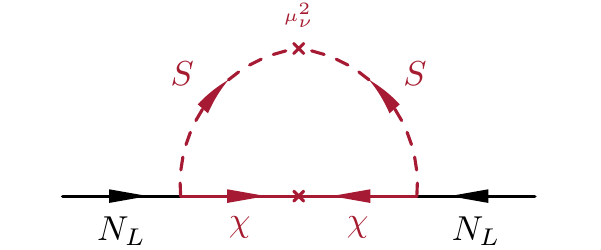}
\end{center}
\caption{Radiative inverse seesaw with DM.}
\label{fig:loop}
\end{figure}

The neutrino mixing matrix in the basis $N^T=(\nu_L, N_{R}^c, N_{L}, \chi^c)$%
, the neutral fermions mass term of the form $\frac{1}{2}\overline{N^c}
\mathcal{M}N +\mathrm{H.c.}$ is:
\begin{equation}
\mathcal{M}=%
\begin{pmatrix}
0 & m_{D}^{\top } & 0 & 0 \\
m_{D} & \epsilon_R & M & 0 \\
0 & M^{\top } & \epsilon_L & 0 \\
0 & 0 & 0 & m_{\chi }%
\end{pmatrix}%
\,, \label{eq:numat}
\end{equation}
where $m_{D}=y_{\nu }\left\langle H\right\rangle $ is the usual Dirac
neutrino mass.

The terms $\epsilon_{L,R}$ are generated radiatively in this model; the loop
contribution is made possible thanks to the presence of the coupling $%
y_{N}N_{L}\bar{\chi}S$ in \eq{lag}, the soft breaking term in
\eq{pot} which allows for the chirality flip, and the
condition $\left\langle S\right\rangle =0$ which forbids the
tree-level contribution. In the tree-level limit, i.e.,
$\epsilon_{R,L} = 0$, we have three strictly massless neutrinos and
three heavy Dirac pairs of neutrinos -- lepton number is a good
symmetry of the Lagrangian. However, by turning on the terms
$\epsilon_{R,L}$, the additive conservation of lepton number gets
violated and induces small neutrino masses:
\begin{equation}
m_{\nu }\simeq m_{D}^{\top }\frac{1}{M^{\top }}\epsilon_L \frac{1}{M}%
m_{D}\equiv m_{D}^{\top }\mathcal{M}_{R}^{-1}m_{D}\,, \label{eq:mnu}
\end{equation}
at lowest order in $\epsilon_{R,L}$. The neutrino mass contribution from $%
\epsilon_R$ gets an additional loop suppression with respect to that of $%
\epsilon_L$ ~\cite{Dev:2012sg} and so we will ignore it here. The light
neutrino masses are linearly proportional to $\epsilon_L$, which is the term
responsible of lepton number violation. Based on this, one can argue that $%
\epsilon_L$ should be small because in its absence the symmetry of
the theory is enhanced; this is `technical naturalness' in the 't
Hooft sense. Whereas in most models invoking the inverse seesaw
mechanism to generate neutrino masses $\epsilon_L$ is \emph{assumed} to be
tiny, here this is justified by the fact that it is generated radiatively via a loop
which involves our DM candidate.

The matrix $\epsilon_L$ is induced by the diagrams in \fig{loop}, and is found to be:
\begin{eqnarray}
\epsilon_L &=&-i\frac{y_{N}^{2}\mu _{\nu }^{2}m_{\chi }}{32\pi ^{2}}\left[ {%
\frac{\left( m_{S}^{2}+3m_{\chi }^{2}\right) }{\left( m_{\chi
}^{2}-m_{S}^{2}\right) ^{2}}+}\frac{m_{\chi }^{2}\left( 3m_{S}^{2}+m_{\chi
}^{2}\right) }{\left( m_{\chi }^{2}-m_{S}^{2}\right) ^{3}}\log \frac{%
m_{S}^{2}}{m_{\chi }^{2}}\right] \nn \\
&\simeq &\frac{y_{N}^{2}}{32\pi ^{2}}\frac{m_{\chi
}}{m_{S}^{2}}\,\mu _{\nu }^{2}~\quad\quad\text{for }m_{\chi }\ll
m_{S}\,,
\end{eqnarray}
%
We can get naturally small values for $\epsilon_L$ without having to
put by hand a number which is far from the weak scale in the Lagrangian. For instance, for $\mu_\nu
= 10 \gev$, $m_\chi =100 \gev$, and $m_S = 1 \tev$, we get $\epsilon_L\approx 3 \kev$. With the additional freedom introduced by $\epsilon_L$,
neutrino masses become decoupled from the mixing between light and heavy
neutrinos and therefore from the strength of lepton flavor violation~\cite%
{Bernabeu:1987gr}. This makes such a framework particularly rich
phenomenologically, as it leads to many signals at low energy physics experiments (see, e.g., \cite{Malinsky:2009df,Forero:2011pc,LalAwasthi:2011aa,Antusch:2014woa})
 as well as high energy colliders (e.g., \cite{Deppisch:2005zm,BhupalDev:2012zg,Das:2012ze,Deppisch:2013cya,Das:2014jxa,Das:2015toa,Antusch:2015mia}).

Neutrino masses and mixing angles can be accommodated for a given
$\epsilon_L$, that is for a set of parameters \{$m_S, m_\chi, y_N,
\mu_\nu$\}, by using the freedom we have on the neutrino Yukawa
coupling. Assuming $\mathcal{M}_{R}$ to be diagonal, the Yukawa
couplings appearing in $m_{D}$ can be parameterized
as~\cite{Casas:2001sr}:
\begin{equation}
y_{\nu }=\sqrt{\mathcal{M}_{R}}\,\mathcal{R}\,\sqrt{\hat{m}_{\nu }}
U_{lep}^{\dagger }\,, \label{eq:yuk}
\end{equation}
where $U_{lep}$ is the lepton mixing matrix, $U_{lep}^{\top }\,m_{\nu
}\,U_{lep}=\mathrm{diag}(m_{1},m_{2},m_{3})\equiv \hat{m}_{\nu }\left\langle
H\right\rangle ^{2}$, and $\mathcal{R}$ is an orthogonal matrix.
Because of this freedom, we find that the neutrino parameters as well as
limits on unitarity deviations and flavor changing currents can easily be
accommodated.

\section{Higgs physics and Electroweak Phase Transition}
\label{sec:ewpt}
In this section, we will discuss different issues related to the scalar
sector such as the radiative corrections to the Higgs mass, the Higgs
invisible decay and the electroweak phase transition strength.

\subsection*{Higgs Mass}

\begin{figure}[t!]
\begin{center}
\includegraphics[width=0.5\textwidth]{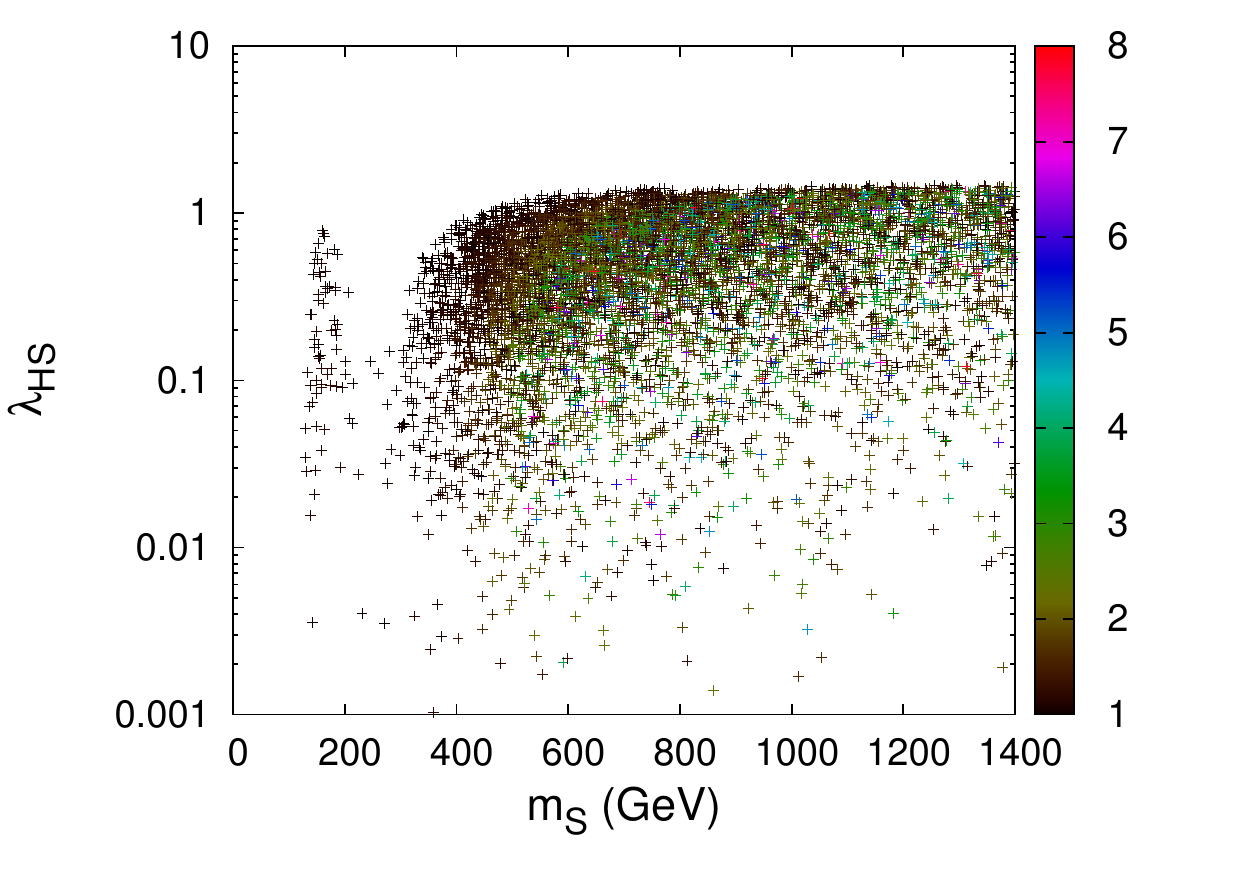}
\end{center}
\caption{The coupling $\lambda_{HS}$\ in absolute value versus the
scalar mass, and the palette reads the phase transition strength, eq.~(\ref{eq:v/t}).}
\label{fig:vctc}
\end{figure}

The Higgs mass at one-loop can be estimated as the second derivative of the
zero-temperature effective potential that is given in the $\overline{DR}%
^{\prime }$ scheme by~\cite{Martin:2001vx}:
\begin{eqnarray}
V_{1-l}^{T=0}\left( h\right) &=&-\frac{\mu
_{H}^{2}}{2}h^{2}+\frac{\lambda_{H}}{8}h^{4}\nn\\
&&+\sum_{i}n_{i}\frac{m_{i}^{4}(h)}{64\pi
^{2}}\left( \log \frac{m_{i}^{2}(h)}{\Lambda
^{2}}-\frac{3}{2}\right)\,,
\end{eqnarray}
where $h=(\sqrt{2}\Re(H^{0})-\upsilon )$\ is the real part of the neutral
component in the doublet, $n_{i}$\ are the field multiplicity, $\Lambda $ is
the renormalization scale which we choose to be the Higgs mass, 125 GeV, and
$m_{i}^{2}(h)$ are the field-dependent mass squared that can be written as $%
m_{i}^{2}(h)=\alpha _{i}+\frac{1}{2}\beta _{i}h^{2}$, i.e.,
\begin{align}
m_{W}^{2}(h)& =\tfrac{1}{4}g^{2}h^{2} \nn \\
m_{Z}^{2}(h)&=\tfrac{1}{4}\left(g^{2}+g^{\prime 2}\right) h^{2}\nn \\
m_{t}^{2}(h)&=\tfrac{1}{2}y_{t}^{2}h^{2} \nn\\
m_{h}^{2}(h)& =-\mu _{H}^{2}+\tfrac{3}{2}\lambda _{H}h^{2}\nn\\
m_{G}^{2}(h)&=-\mu _{H}^{2}+\tfrac{1}{2}\lambda _{H}h^{2} \nn \\
m_{S_{r,i}}^{2}(h)& =\mu _{S}^{2}+\tfrac{1}{2}\lambda _{HS}h^{2}\pm \tfrac{1}{2}\mu _{\nu }^{2}\,,
\end{align}
where $G$\ and $S_{r}\equiv \Re (S)$ ($S_{i}\equiv \Im (S)$) denote the
Goldstone bosons and the real (imaginary) part of the complex scalar $S$,
respectively. In all what follows we will consider the case $\mu _{S}^{2}+\tfrac{1}{2}\lambda _{HS}\upsilon^{2} \gg  \tfrac{1}{2}\mu _{\nu }^{2}$, such that
$m_S \equiv m_{S_r}\sim m_{S_i}$.\\

The term $\mu _{H}^{2}$\ can be eliminated in favor of the doublet vacuum expectation value (vev) via
the tadpole condition at one-loop:
\begin{equation}
\mu _{H}^{2}=\tfrac{1}{2}\lambda _{H}v^{2}+\frac{1}{32\pi ^{2}}%
\sum_{i}n_{i}\beta _{i}m_{i}^{2}\left( \log \frac{m_{i}^{2}}{\Lambda ^{2}}%
-1\right) \,,
\label{eq:muH}
\end{equation}
with $\dot{m}_{i}^{2}(h)=\partial m_{i}^{2}(h)/\partial h$\ and by using \eq{muH}, the Higgs mass squared can be written as:
\begin{equation}
m_{h}^{2}=\lambda _{H}v^{2}+\frac{v^{2}}{32\pi
^{2}}\sum_{i}n_{i}\beta _{i}^{2}\log \frac{m_{i}^{2}}{\Lambda ^{2}}\,.
\label{eq:Mh}
\end{equation}

In order to explain the discovered scalar resonance at
$m_{h}=125.09\mp 0.21$ GeV~\cite{Aad:2014aba, CMS:2014ega},
the Higgs quartic coupling $\lambda_{H}$ has to be adjusted according to the radiative corrections.

\subsection*{Higgs Invisible Decay}

\begin{figure}[t!]
\begin{center}
\includegraphics[width=0.2\textwidth]{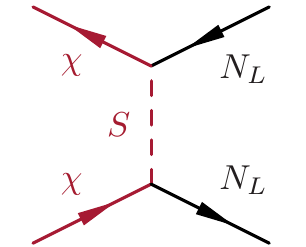}~~~\includegraphics[width=0.2\textwidth]{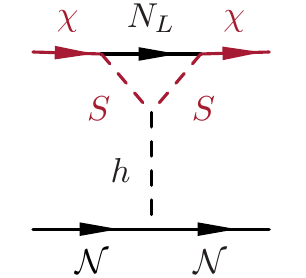}
\end{center}
\caption{DM production (left) and direct detection (right) diagrams.}
\label{fig:dmproddd}
\end{figure}

If either of $m_S$ or $m_\chi$ is smaller than half of the Higgs
mass then the bounds on Higgs invisible decay should be fulfilled,
i.e., $\mathcal{B}(h\rightarrow inv)<17\%$ ~\cite{Bechtle:2014ewa}.
Here, the Higgs invisible decay branching ratio is given by:
\begin{equation}
\mathcal{B}(h\rightarrow inv)=\frac{\sum\nolimits_{X}\Gamma \left(
h\rightarrow XX^{c}\right) }{\sum\nolimits_{X}\Gamma \left( h\rightarrow
XX^{c}\right) +\Gamma _{\mathrm{SM}}},
\end{equation}
where $X\equiv \chi ,S_{r},S_{i}$ and $\Gamma _{\mathrm{SM}}=4.434~\mathrm{%
MeV}$\ is the SM Higgs decay width. The Higgs decay widths to $S$ and $\chi$  are given by:
\begin{align}
\Gamma \left( h\rightarrow SS^\star\right) & =\frac{\lambda _{HS}^{2}\upsilon ^{2}%
}{16\pi m_{h}}\,\left( 1-\frac{4m_{S}^{2}}{m_{h}^{2}}\right) ^{\frac{1}{2}%
}\Theta \left( m_{h}-2m_{S}\right) ,  \label{Br1} \\
\Gamma \left( h\rightarrow \chi \chi ^{c}\right) & =\frac{y_{\chi }^{2}}{%
16\pi }m_{h}\left( 1-\frac{4m_{\chi }^{2}}{m_{h}^{2}}\right) ^{\frac{3}{2}%
}\Theta \left( m_{h}-2m_{\chi }\right)\,.  \label{Br2}
\end{align}
The decay to $\chi$ occurs through the triangle one-loop vertex $y_{\chi }h%
\bar{\chi}\chi $, that is shown in \fig{dmproddd}. This effective
vertex is given in terms of the parameters of the model by:
\begin{equation}
y_{\chi }=\frac{\lambda _{HS}v}{16\pi ^{2}m_{S}^{2}}%
\sum_{i}y_{N_{i}}^{2}M_{i}\,Q\left( \frac{M_{i}^{2}}{m_{S}^{2}}\right) \,,
\label{eq:hxx}
\end{equation}
where the function $Q$ is defined as:
\begin{equation}
Q\left( x\right) =
\begin{cases}
\begin{array}{ccc}
Q^+(x)  &  & x>\frac{1}{4} \\[3mm]
Q^-(x)  &  & x<\frac{1}{4}\,.%
\end{array}%
\end{cases}%
%
\end{equation}
with
\begin{eqnarray}
Q^+(x)&=&\frac{2}{\sqrt{y}}\left[ \arctan \left( \frac{2x-1}{\sqrt{y}}\right)
+\arctan \left( \frac{1}{\sqrt{y}}\right) \right]\nn\\
Q^-(x)&=&\frac{1}{\sqrt{-y}}\left( \log \left( \frac{2x-1-\sqrt{-y}}{2x-1+\sqrt{-y}}\right) -\log \left( \frac{1+\sqrt{-y}}{1-\sqrt{-y}}\right)
\right)\nn
\end{eqnarray}
where $y=4x-1$.


When only the $S$ channel is open, this constraint can be translated as an upper bound on the quartic coupling:
\begin{equation}
\lambda_{HS} \left( 1-\frac{4m_{S}^{2}}{m_{h}^{2}}%
\right) ^{1/4}\lesssim9.7\times\,10^{-3}.\,  \label{eq:InvCons}
\end{equation}
\subsection*{Electroweak Phase Transition}

\begin{figure}[t!]
\includegraphics[width=0.5\textwidth]{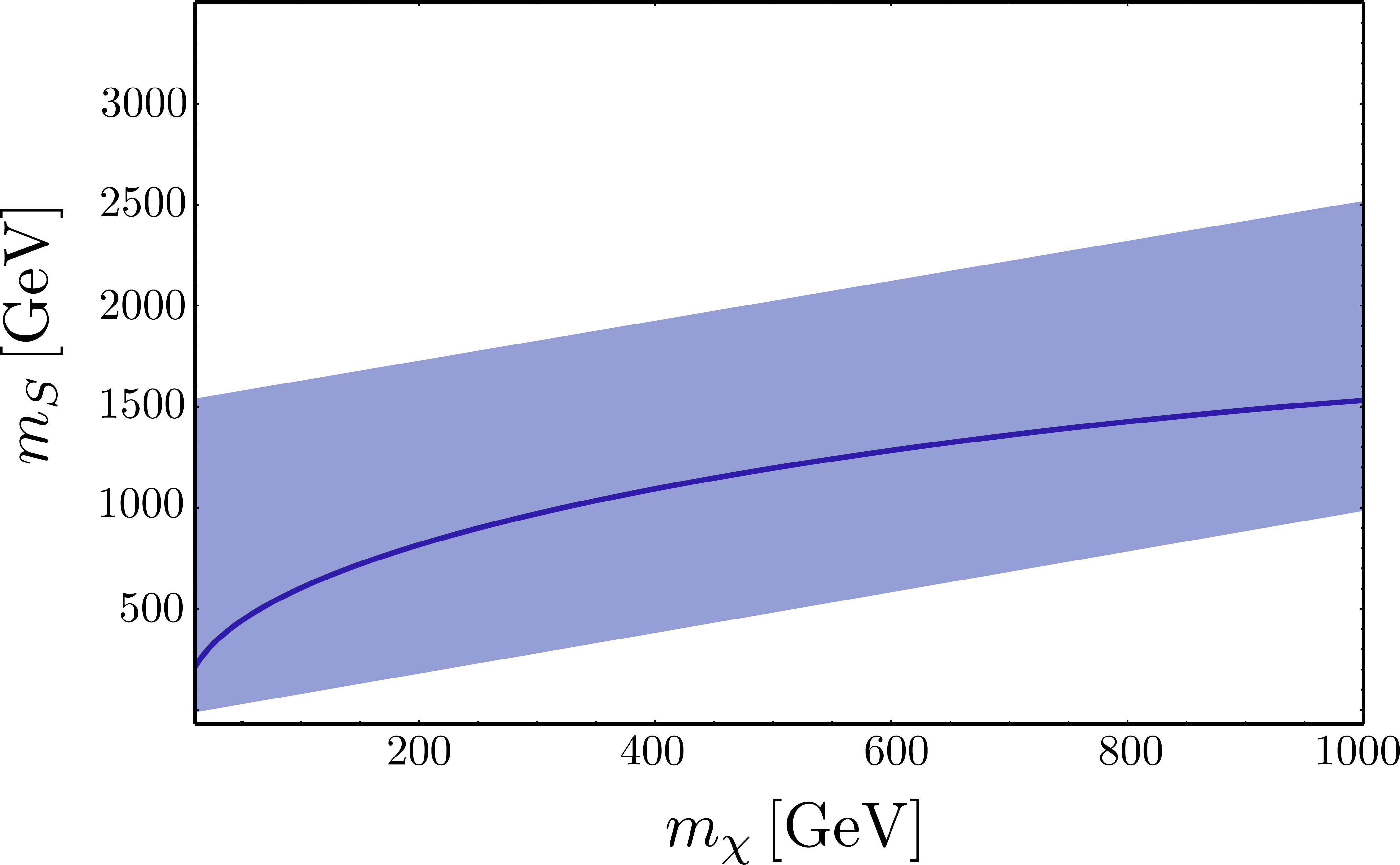}
\caption{The mediator mass as a function of the DM mass. All the
benchmark points satisfy the relic density abundance and the detection
limits. The solid line is the particular benchmark defined by $y_{N}=1$ and $M_{1}=10$
GeV.}
\label{fig:omega-l}
\end{figure}

The SM has all the qualitative ingredients for electroweak
baryogenesis, however the generated matter-antimatter asymmetry cannot account for observations
due the smallness of $CP$ violation and the fact that electroweak
phase transition (EWPT) is not strongly first order~\cite{Trodden:1998ym}. However, it is well
known that the EWPT can be strengthened if new scalar degrees of
freedom coupled to the Higgs are added~\cite{Anderson:1991zb,Espinosa:1993bs,Choi:1993cv,McDonald:1993ey,Ham:2004cf,Ahriche:2007jp, Espinosa:2007qk,Profumo:2007wc}.
In this model, we have two extra scalar degrees of freedom around the weak scale, so we expect an enhancement of the EWPT strength.\\

In order to investigate the nature of the EWPT, the effective
potential should be properly defined at finite temperature. The full
effective potential can be written as \cite{Dolan:1973qd,Weinberg:1974hy}:
%
\begin{eqnarray}
V_{eff}(h,T)&=&V_{1-l}^{T=0}\left( h\right) +\frac{T^{4}}{2\pi ^{2}}\sum_{i}n_{i} \,J (\tfrac{m_i(h)^2}{T^2}),\\
J(\beta)&=&\int_{0}^{\infty } x\log \left( 1 + \eta\, \exp^{-\sqrt{x^{2}+\beta}}\right)dx\nn \,,
\label{eq:Veff}
\end{eqnarray}

with $\eta= -1\, (+1)$ for bosons (fermions). Another bosonic thermal
contribution should be included in \eq{Veff}. This contribution represents
a leading part of higher order corrections that is estimated by performing
the resummation of an infinite class of infrared-divergent multi-loops,
known as the ring (or daisy) diagrams~\cite{Carrington:1991hz}. We will include this effect by replacing the field-dependent masses of
the scalar and longitudinal degrees of freedom by their thermal corrections
$\tilde{m}_{i}^{2}(h,T)=m_{i}^{2}(h)+\Pi _{i}(T)$, where the thermal parts
are given by:
\begin{align}
\Pi _{W}^{L}& =\frac{11}{6}g^{2}T^{2}\,,
\quad\Pi _{W}^{\top }=\Pi _{Z}^{\top }=0\,,\quad~\Pi _{B}=\frac{11}{16}g^{{\prime }2}T^{2}\,,\quad \nn \\
%
\Pi _{h}& =\Pi _{\chi }=\left( \tfrac{1}{2}\lambda _{H}+\tfrac{1}{6}\lambda_{HS}+\tfrac{1}{4}g^{2}+\tfrac{1}{12}g^{\prime 2}+\tfrac{1}{4}%
y_{t}^{2}\right) T^{2}\,,\nn \\
\quad\Pi _{S}&=\left( \tfrac{1}{3}\lambda _{S}+\tfrac{1}{3}%
\lambda_{HS}\right) T^{2}\,,
\end{align}
with $g$, $g^{\prime }$ and $y_{t}$ are the gauge and Yukawa couplings.\\

To generate a net baryon asymmetry at the electroweak scale~\cite{Kuzmin:1985mm},
the anomalous interactions that violate $B+L$ should be switched-off inside the nucleated bubbles, which
leads to the strong first order phase transition criterion \cite{Shaposhnikov:1987tw}
\begin{equation}
\upsilon _{c}/T_{c}>1\,, \label{eq:v/t}
\end{equation}
where $T_{c}$ is the critical temperature at which the effective potential
exhibits two degenerate minima, and $\upsilon _{c}$ is the Higgs doublet
vev at this temperature. The transition can be defined by two conditions:
\begin{equation}
\left. \frac{\partial }{\partial h}V_{eff}(h,T_{c})\right\vert
_{h=v_{c}}=0\,,\quad ~V_{eff}(v_{c},T_{c})=V_{eff}(0,T_{c}).
\end{equation}

In the SM, the criterion $\upsilon _{c}/T_{c}$ $\sim \left(
2m_{W}^{3}+m_{Z}^{3}\right) /\left( \lambda _{H}\upsilon ^{3}\right)
>1$ implies that the Higgs mass has to be $m_{h}<42~\mathrm{GeV}$
\cite{Bochkarev:1987wf} in contradiction with measurements. Therefore
in the SM, the electroweak phase transition is a smooth crossover.
However, if the radiative contributions in
the Higgs mass (second term in LHS of \eq{Mh}) are significant,
the doublet quartic coupling $\lambda _{H}$ gets smaller and the
phase transition gets stronger. We perform a random scan in the
parameter space  with 6000 benchmark points taking into
account the Higgs mass and its branching ratio to invisibles. Our results are shown in \fig{vctc}.\\

It is clear from \fig{vctc}
that for the phase transition to be naturally strong, the mixed quartic coupling must satisfy $\lambda _{HS} \gtrsim 10^{-2}$,
which means, after using \eq{InvCons}, that the invisible Higgs decay channel---barring
tuning of parameters---must be closed, i.e., the scalar $S$ mass should be
larger than $m_{h}/2$. One also remarks that, in any case, a strong phase transition favors heavy scalars. Similar behavior had
been seen in \cite{Ahriche:2007jp, Ahriche:2010ny, Ahriche:2012ei,
Ahriche:2013zwa, Ahriche:2015mea, Ahriche:2015taa}, where extra
scalars can help bring about a strongly first-order EWPT by:
(a)~reducing the Higgs quartic coupling $\lambda_{H}$ to small values
and having significant radiative correction to get the correct Higgs mass, \eq{Mh};
%
and (b)~enhancing the value of the effective potential
at the wrong vacuum at the critical temperature without suppressing
the ratio $v_{c}/T_{c}$, which relaxes the severe bound on the mass
of the SM Higgs.\\

At the International Linear Collider (ILC), the triple Higgs coupling $\lambda_{hhh}$ can be
measured with about 20\% accuracy or better at $\sqrt{s}=500$
\textrm{GeV} with integrated luminosity $\mathcal{L}=500$ $fb^{-1}$
\cite{ILC}. Unfortunately, within our numerical scan, one remarks
that the relative enhancement in the triple Higgs model with respect
to the SM,
\begin{equation}
\Delta =\frac{\lambda _{hhh}-\lambda _{hhh}^{SM}}{\lambda _{hhh}^{SM}},
\end{equation}
lies between -2.3\% and 10\%.\\

One has to notice that an extra $CP$-violating source is required to
have a realistic electroweak baryogenesis scenario. Therefore a
$CP$-violating phase should be added in the Lagrangian of the complete
theory. In analogy to a scenario of electroweak baryogenesis from a
singlet scalar, one can modify the top quark
Lagrangian mass term by adding a non-renormalizable dimension 6 operator
where the complex scalar couples to the top-quark, for instance~\cite{Cline:2012hg}
\begin{equation}
\frac{e^{i\alpha}}{\Lambda^2} Q_L H t_R S^2\,.
\end{equation}
This new
interaction is suppressed by a new-physics scale that can be well
above one TeV.

\section{Relic density abundance and direct detection \label{sec:dm}}

The DM candidate can be either the fermion, $\chi $ or the lightest of the
scalars $\Re (S)$ and $\Im (S)$ depending on the sign of $\mu _{\nu }$. The
fermionic case has more predictive power because there are fewer production
channels and the parameters entering the evaluation of the relic density are
directly related to those entering in the neutrino masses. It is conceptually
more elegant than the scalar cases whose phenomenology would be dependent
upon the Higgs portal with no direct relation with neutrino masses. For these
reasons we consider our dark matter candidate to be $\chi $.

In the non-relativistic limit, the thermally averaged annihilation
cross-section can be written as $\left\langle \sigma v_{r}\right\rangle
=a+bv_{r}^{2}$, where $v_{r}\simeq \sqrt{6/x_{f}}$ is the relative DM
velocity and $a$ and $b$ are respectively the $s$-wave and $p$-wave factors
which receive contributions from different annihilation channels.
The relic density is then given by
\begin{equation}
\Omega h^{2}\simeq \frac{1.04\times 10^{9}\mathrm{GeV}^{-1}}{M_{Pl}}\frac{%
x_{f}}{\sqrt{g_{\ast }(T_{f})}\left( a+3b/x_{f}\right) },
\label{eq:DM_Omega}
\end{equation}
where $T_{f}=m_{\chi }/x_{f}$ is the freeze-out temperature, $g_{\ast }(T)$
is the number of relativistic species at temperature $T$, and $M_{Pl}$ is
the Planck mass.

The dark matter production is thermal via the freeze-out mechanism, and
proceeds via the annihilation diagrams depicted in \fig{dmproddd}.
This implies that at least one of the heavy neutrinos is lighter than the
DM. We will consider the hierarchies $M_{1}= M_{2}\equiv M_N \leq m_\chi$ and $M_{3}
> 1$ TeV. $M_i$ are the diagonal entries of the matrix $M$ appearing in \eq{lag}. This hierarchy is found to satisfy the neutrino oscillation data
and flavor-changing limits for any mass $m_\chi$ in the range GeV-TeV.

In the limit $ m_S\gg m_{\chi }\gg M_{N}$, the annihilation cross section $%
\chi \chi \to N_{L_i} N_{L_j}$ is given by
\begin{equation}
\left\langle \sigma v\right\rangle \simeq \frac{y_{N_i}^{2}
y_{N_j}^{2} m_{\chi }^{2}}{48\pi m_{S}^{4}}\,v_{r}^{2}\,,
\label{eq:sv}
\end{equation}
and using, \eq{sv} and \eq{DM_Omega}, the good relic abundance is obtained for
\begin{equation}
\frac{\Omega h^2}{0.12} \simeq \left(\frac{x_f}{20}\right) \left(\frac{1}{y_N}%
\right)^4 \left( \frac{500 \,\mathrm{GeV}}{m_\chi}\right)^2 \left( \frac{m_S%
}{\mathrm{TeV}} \right)^4 \,. \label{eq:approxO}
\end{equation}

Therefore, we see that the model quite naturally reproduces the
relic density constraint. In order to cover the full parameter
space, we implemented the model in \texttt{micrOmegas}
\cite{Belanger:2014vza}.\footnote{We used the \texttt{FeynRules}
\cite{Alloul:2013bka} package to generate the model file.}  We perform a random scan on the relevant
parameters of the model in the following ranges:
\begin{eqnarray}
10\,\mathrm{GeV} &\leq &m_{\chi }\leq 1\,\mathrm{TeV} \notag
\label{eq:range} \\
1\,\mathrm{GeV} &\leq &M_N\leq 1.05\,m_{\chi } \notag \\
1\,\mathrm{TeV} &\leq &M_{3}\leq 2.5\,\mathrm{TeV} \\
 m_h/2 &< &m_{S}\leq 1.5\,\mathrm{TeV} \notag \\
10^{-4} &\leq &|y_{N_{i}}|,\lambda _{HS}\leq \sqrt{4\pi }\,, \notag
\end{eqnarray}
and we fix $\mu _{\nu }=1\,\mathrm{GeV}$. For simplicity we take $%
y_{N_{1}}=y_{N_{2}}\equiv y_{N}$.\\

\begin{figure}[t!]
\begin{center}
\includegraphics[width=0.5\textwidth]{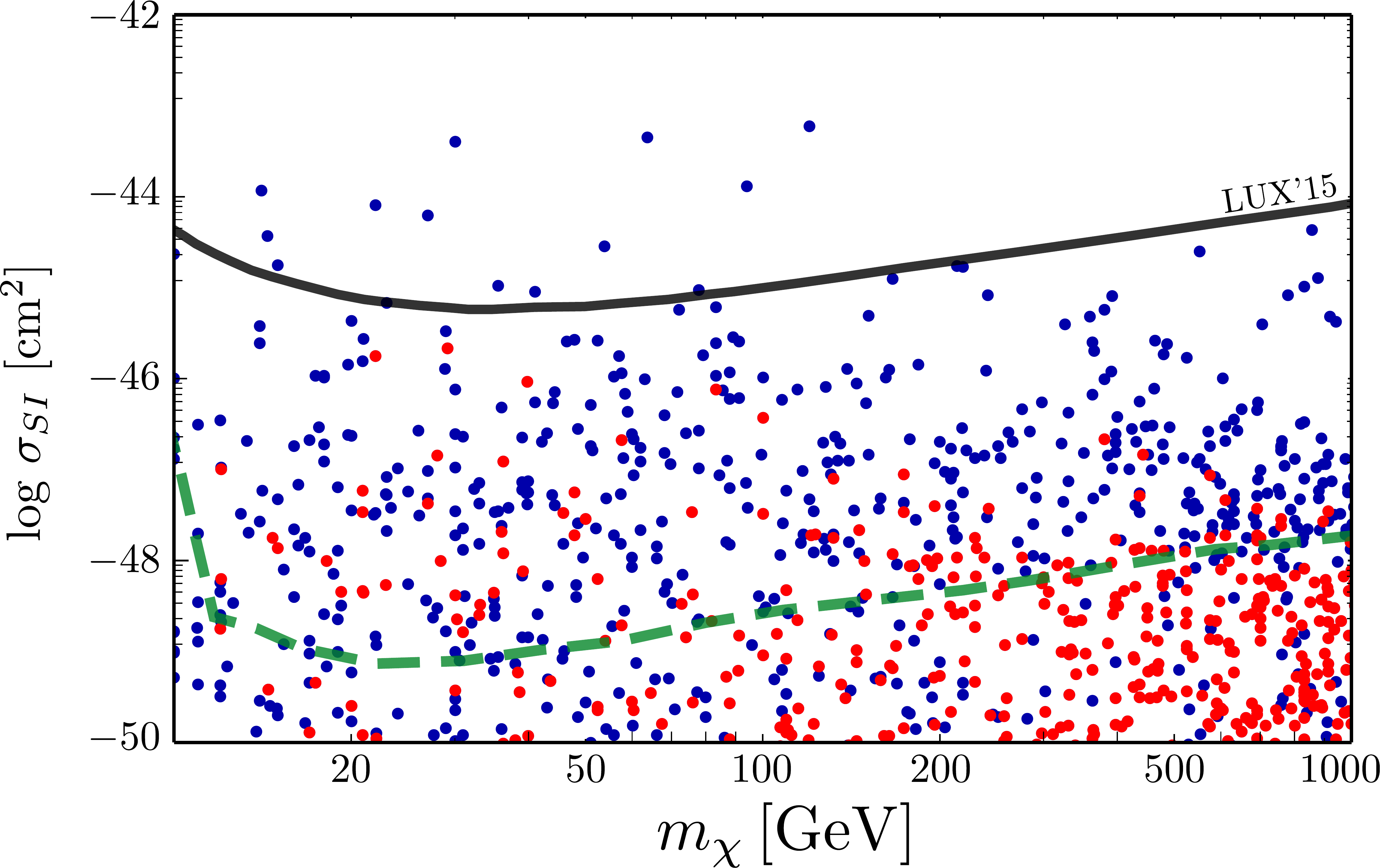}
\caption{Direct detection cross section versus the DM mass. Solid line is the limit from LUX experiment~\cite{Akerib:2015rjg}
and the dashed line is the neutrino floor. The points in red satisfy the conservative bound on the couplings $|y_{N_i}|, \lambda_{HS}<1$.}
\label{fig:omega-r}
\end{center}
\end{figure}

In \fig{omega-l}, we show the allowed parameter space in the plane $m_{S}$ versus DM mass, $m_\chi$.
The solid line represents the evolution of the particular benchmark model
defined by $y_{N}=1$ and $M_{1}=10$ GeV, which confirms the estimate in \eq{approxO}.
All the points in \fig{omega-l} and \fig{omega-r} satisfy the relic density constraint
\cite{Ade:2015xua}
\begin{equation}
\Omega h^{2}=0.1198\pm 0.0015\,,
\end{equation}
as well as the requirement for a strong first order phase transition, \eq{v/t}, and the bound on Higgs invisible decay.\\

The direct detection is obtained via the radiative diagram shown in \fig{dmproddd}.
We have implemented the effective interaction vertex, \eq{hxx}, in our model.
%
%
%
In \fig{omega-r}, we show the expected spin-independent cross
section as a function of the DM mass for the range of parameters defined in
\eq{range}. We show also the points which satisfy a more conservative bound on the couplings, $|y_{N_i}|, \lambda_{HS}<1$. The solid line is the current best limit on direct
detection experiments, obtained from the latest results of the LUX
experiments~\cite{Akerib:2015rjg} which are the strongest to date. The dashed line illustrates the `neutrino floor'. As
apparent from the plot, the model is already probed by direct detection
experiments even thought the scattering is loop-suppressed, and future runs
of current experiments as well as planned experiments will probe a
significant portion of the parameter space of the model.

\section{Conclusions\label{sec:end}}

In this paper, we have presented a model which provides simultaneous explanations
for neutrino masses and dark matter.
Neutrino oscillations are accounted for thanks to a DM-assisted radiative inverse seesaw
mechanism, where  the small lepton number violating parameter is generated at the one-loop level.
There is no assumed hierarchy in the mass scales of the model and they can all be $\order{\gev-\tev}$.
The symmetry which precludes the tree-level inverse seesaw contribution provides at the same time a
fermionic dark matter candidate  whose abundance is consistent with cosmological data and its scattering cross section off nuclei satisfies the latest LUX bound and can be probed by future DM direct detection experiments.
The scalar responsible of the DM interactions with the visible sector as well as the generation of neutrino masses triggers a strong
enough electroweak phase transition to make electroweak baryogenesis viable.

\bigskip
\bigskip
\footnotesize

{\bf Acknowledgment}
AA is supported by the Algerian Ministry of Higher Education and
Scientific Research under the CNEPRU Project No. D01720130042. SMB
thanks F. Deppisch for discussions and acknowledges financial
support from the research grant "Theoretical Astroparticle Physics"
number 2012CPPYP7 under the program PRIN 2012 funded by the Italian
MIUR and from the INFN ``IS'' Theoretical Astroparticle Physics
(TAsP-LNF) and support of the spanish MICINN's Consolider-Ingenio
2010 Programme under grant MultiDark CSD2009-00064.

\bigskip
\bigskip
\bibliography{refs,merged,extraref}

\end{document}